\documentstyle[preprint,aps]{revtex}
\begin{document}

\title{Stability of Designed Proteins Against Mutations}

\author{R. A. Broglia$^{1,2,3}$, G. Tiana$^{3,1}$, H. E. Roman$^{1,2}$, 
        E. Vigezzi$^{1,2}$ and  E. Shakhnovich$^4$ }

\address{$^1$Dipartimento di Fisica, Universit\`a di Milano,
         Via Celoria 16, I-20133 Milano, Italy.}

\address{$^2$INFN, Sezione di Milano, Via Celoria 16, I-20133 Milano, Italy.}

\address{$^3$The Niels Bohr Institute, University of Copenhagen,
         2100 Copenhagen, Denmark.}

\address{$^4$Department of Chemistry and Chemical Biology, Harvard University, 12 Oxford Street,
         Cambridge, MA 02138}

\date{\today}

\maketitle

\smallskip

\begin{abstract}
The stability of model proteins with designed sequences 
is assessed in terms of the
 number of sequences (obtained from the designed sequence through mutations),
 which fold into the ``native'' conformation. By a complete enumeration of the
 total number of sequences obtained by introducing up to 4 point 
mutations and  up to 7 composition--conserving mutations (swapping of amino acids) in a 
 36mers chain, it
 is found that there are $10^8-10^9$ sequences which in the folding process
 target onto the ``native'' conformation. Consequently, 
 proteins with designed sequences display a remarkable degree
 of stability and, to a large extent, of designability.
\end{abstract}
\bigskip

\newpage
\narrowtext
A number of previous analyses \cite{GSW,SSK1,PNAS,PRLF,GOLDSTEIN,LENGTH}
(reviewed in \cite{BRYN_REVIEW,COSB}) have provided arguments and 
supporting evidence
for the deep connection existing between the energetic properties 
of protein sequences and their ability to fold fast into
their native conformations. In particular it was
found \cite{SSK1,PNAS,PRLF} that the presence of a large (compared to the dispersion
of interaction energies) energy gap between
the native state and  the bulk of misfolded conformations
that are structurally dissimilar to the native state
is an important factor that ensures fast folding
into the native conformation (foldability requirement). 
 A number of observations
support the notion that sequences of natural proteins
have been optimized to satisfy the foldability requirement:

1) Random sequences undergo non-cooperative folding transition
\cite{Goldberg_F2,SAUER} while designed sequences and
proteins fold cooperatively \cite{GSW,PNAS,PRLF,Privalov_APC,CI2_0}.

2) The native state of random sequences is very unstable even
to small changes in potential function \cite{bryngelson}
while the ones that have larger gaps are much more
robust
with respect to changes in the energy function
\cite{PGT2,Antonio_96}. The latter behavior
is characteristic  of real proteins that exhibit the remarkable
ability to maintain their native structure intact 
in a wide range of conditions including variation of
temperature, pH, solvent composition etc. 

3) It was shown theoretically that ground (native) states
of random sequences are very unstable with respect to point mutations:
the probability that a mutated sequence has the same native state
scales as $\gamma^{-8}$ where $\gamma$ is the number of conformations
per one residue in the chain \cite{POINTMUT}.
In contrast,  real proteins are able to accomodate
numerous mutations that are neutral with respect to structure
changes \cite{CREIGHT} . This fact has obvious implication
for the molecular evolution of proteins: it accounts for the existence
of large families of proteins that may have 
diverged from a common root. Proteins belonging 
to a family have
homologous sequences and their
native states are structurally similar. 

While the stability
of designed sequences with respect to point mutations
has been demonstrated \cite{Tiana_98} in simulations,
and the fact that larger energy gaps imply greater ability
of the designed sequence to accomodate many neutral mutations
acknowledged \cite{GSW,SSK1,PNAS,PRLF,GoldsteinPNAS,DES_REV,Michele_BI},
the actual quantitative analysis of how many mutations exist 
that preserve the native state was missing.

In what follows we present a quantitative
analysis of how many neutral mutations can  proteinlike 
sequences of various degree of gap optimization  
accomodate. The outcome of this analysis is baffling. In fact, it 
will be concluded that designed proteins can accomodate billions
of multiple mutations without changing their ability to fold on
short call into the native conformation.

For the analysis we use lattice model of a protein that
has been used earlier by us \cite{Tiana_98,NUCLEUS,TRAPS}
and others \cite{SO1,KT}. The model sequences are composed
of aminoacids of 20 types and contain 36 monomers. Two aminoacids
are considered interacting if they occupy neighboring 
positions on the lattice but are not sequence neighbors.
The energy of the interaction depends on the identity of the
aminoacids involved, so that there is a $20\times 20$ parameter
matrix that describes the energetics in the model. We used the
set of parameters suggested by Miyazawa and Jernigan
( table 6 of Ref. \cite{MJ}).
The associated standard deviation of the interaction energies
between different aminoacid types is
$\sigma=0.3$.

Our approach  to protein simulations is based on the idea
of designing sequences having a large energy gap in the target
conformation chosen to serve as a native state for simulations
\cite{PRLF,ALAMUT}. 

A sequence that  has sufficiently low energy in a 
conformation chosen as native is denoted as S$_{36}$ 
(cf. caption to Fig. 1). This sequence
is the same as
was studied in previous publications \cite{Tiana_98,NUCLEUS,TRAPS}. 
In the units we are
considering ($RT_{\rm room}=0.6$ kcal/mol), the energy of S$_{36}$ in its 
native conformation (cf. Fig.~1(a) ) is $E_{\rm nat}=-16.5$. 
Starting from a random configuration, the sequence
S$_{36}$ always reaches the native configuration, and it does it
in a rather short time, of the order of $10^6$ MC
steps. This is a consequence of the fact that the
value of the energy gap
$\delta$ (=2.5), that is, the energy difference between the native and the
lowest dissimilar configuration (configuration with a similarity parameter
$q$ \cite{EVOL1} much smaller than one)  is large, much larger 
than the variance of the contact energies.
The goal of our present analysis is to characterize quantitatively
how many mutations can $S_{36}$ tolerate without losing the 
ability to fold into its native state. In other words,
our study aims at providing an estimate of the number of sequences,
having a certain degree of homology to $S_{36}$, that fold into
its native structure.

To characterize quantitatively single or multiple
mutations, we ascribe to them a value 
$\Delta E$ \cite{Tiana_98}, defined as the difference between
the energies of the altered sequence (S'$_{36}$) and of the
intact chain (S$_{36}$), both calculated in the native
configuration (Fig. 1(a) ). The quantity $\Delta E$ is a measure of
how the energy gap changes upon a mutation
provided that the distribution of energies  of 
conformations that are dissimilar to the native state remains
unaffected by the mutation \cite{Tiana_98}. This was shown to be 
the case when mutations
do not change the aminoacid compositions
\cite{PNAS,PRLF,Tiana_98}. 
In this study we have analysed both the kinds of mutations which
conserve and which do not conserve the composition of
the protein. This gives a lower and an upper limit for the number
of mutations which the ``wild--type'' sequence can tollerate. 

A complete enumeration of all sequences S$'_{36}$ has been done up to
seven mutations keeping fixed the amino acid composition of the
chain (swapping), and up to four without this constraint (pointlike). 
Simulating the dynamics \cite{PRLF} of fifty sequences chosen among
the mutated sequences, with the same composition of S$_{36}$ and with 
$\Delta E<\delta$, it turned out that in $100\%$ of the cases, they can 
reach the native conformation in a time comparable to the folding
time of S$_{36}$. Repeating the same analysis on fifty sequences with up
to four pointlike mutations, we observed that only in three cases the chain 
finds conformations dissimilar from the native one, with lower energy, and
it is not able to find the native conformation within the simulation time.

Further we studied the impact of pointlike mutations on sequences having different degree
of design. To this end we calculated the distributions $n_2(\Delta E)$ associated
with two pointlike mutations for the case of three 
sequences designed to fold into the
structure shown in Fig.1(a) 
with different energy gaps. The distributions for all three
sequences appear to be very similar to each other (Fig.2).
 This is also true for 
 composition--conserving mutations and
for the different numbers of mutations we have analyzed (data not shown). 
These results suggest
that the distribution $n_m(\Delta E)$ associated with $m$ mutations has some
degree of universality. 

Given a sequence characterized by
an energy gap $\delta$, it is then possible to calculate the number
of sequences which fold to the same native structure 
(i.e. for which there is still some energy gap between the native structure
and the bulk of decoys) 
and which differ
from the "wild--type" sequence by $m$ mutations.
To do this one has to  calculate the quantity
\begin{equation}
N_m(\delta)=\int_{-\infty}^{\delta} dE\; n_m(E).
\end{equation}
\noindent
As an example we provide 
the function $N_m(\delta)$ for $m=4$ and for the case of 
pointlike mutations (Fig.3). The corresponding values of $N_m$ for the
sequence S$_{36}$ (whose gap is $\delta=2.5$) for up to 
4 pointlike and 7 swap mutations are shown in Table 1.
The calculation of the total number of sequences $N(\delta)=
\sum_{m} N_m(\delta)$ is beyond our calculational power, and can be
established only with approximate methods \cite{desig_2}. The results
obtained with the ``small'' number of mutations shown in Table 1, and 
which provide a lower limit to the total number of sequences folding
to the same native structure is in any case 
impressive, namely $10^{8}-10^9$.
The same study has been repeated using other two fully compact target
structures (Figs. 1(b) and 1(c) ), generated by the collapse of a 36 
monomers homopolymeric chain
at  low temperature (below the $\theta$--point, see e.g.
\cite{lifshitz}). 
The results are virtually identical to the ones shown in Fig. 2.

The present study further suggests that the normalized gap $\xi=\delta/\sigma$
(or the closely related to it z-score \cite{GSW,EVOL1,EISENBERG})
is a major determinant of the ability
of sequences to fold. To this end, the ''resilience'' of sequences
against point mutations is directly related to their energetic
impact: if the cumulative effect of mutations on
the energy of the native state is weak enough so that the energy gap 
 for the native state remains, the mutations are neutral and the 
mutated sequences will still fold into the native state, albeit at a decreased
stability.  Therefore, the whole issue of the estimating the number
of mutations that are tolerated by a sequence (and hence the number
of homologous sequences that fold to the same conformation)
is reduced to enumerating mutations that keep the energy gap as defined
before. 

Another aspect of sequence design known as ''designability''
was discussed by a number of authors \cite{GoldsteinPNAS,FGB,TANG}.
The concept of ''designability'' focuses on the entropy in sequence space
stating that structures that can accomodate more sequences that
have them as the non--degenerate ground state are more
''designable'', and represent the structures of naturally existing proteins. 
Our study is not entirely unrelated to the issue of designability since
it shows that the greater the gap is, the more sequences homologous
to the ``wild type'' sequence exist that target on the ``native''
conformation in the folding process and, as a consequence, has this 
conformation as its non--degenerate ground state.
However, it addresses in fact a different question, namely: how many 
sequences homologous
to the ''wild-type'' exist that fold into the same 
conformation, being this number a lower limit for the 
degree of designability of a structure. In order to
fully
address this question within our approach one has to consider
two further issues: a) that there exist many non-homologous sequences that
can still fold to the same conformation \cite{DES_REV,FSSP,PE},
and b) that the designability principle emphasizes the strong dependence
of the number of sequences on the properties of the target structure.
In any case, we have provided circumstantial evidence concerning the fact 
that designability of a given structure may
be closely related to the maximal gap with which sequences can be
fit into it. (A similar point was also made earlier by several 
authors 
\cite{GoldsteinPNAS,DES_REV,Michele_BI,FGB}). 
In this case the issue of designability reduces to the question
of what structures allow sequences with greater gaps. We are planning
to address this issue in the near future.

Summing up, in this paper we provided a quantitative estimate 
of the number of mutated sequences that are still able to
fold to the same conformation and found it to be
''astronomically'' large.
The actual number of these sequences,
and thus the designability of the corresponding conformation, is
controlled by the dimensionless parameter $\xi=\delta/\sigma$,
which in turn also controls the folding ability of the notional protein.

\section*{Acknowledgements}
This work was partially supported by NIH grant RO1 GM52126 (to ES).
Financial support by NATO under grant CRG 940231 is gratefully 
acknowledged.

\begin{table}
\caption{The number of mutated sequences S'$_{36}$ which fold into the
native conformation shown in Fig 1(a). In column one the number of
mutations $m$ is shown. Columns 2 and 3 are associated with composition
conserving results (c.), while columns 3 and 4 correspond to 
pointlike mutations (n.~c.).
Columns 2 and 4 display the number of sequences associated with a 
change in energy $\Delta E$ smaller than the gap $\delta$, while
columns 3 and 5 display the total number of sequences associated with 
the number of mutations $m$.} 
\smallskip
\begin{tabular}{|c|c|c|c|c|}
\tableline
m & $\Delta E<\delta$ (c.) & Tot (c.) & $\Delta E<\delta$ (n.c.) & Tot
(n.c.) \\\hline
1 &     &     & 613 & 684 \\\hline
2 & 447	& 630 & $1.59\cdot 10^5$ & $2.27\cdot 10^5$ \\\hline
3 & 3339 & 14280 & $2.30\cdot 10^7$ & $4.89\cdot 10^7$ \\\hline
4 & $1.37\cdot 10^5$ & $5.30\cdot 10^5$ & $1.99\cdot 10^9$ & $7.68\cdot 10^9$ \\\hline
5 & $4.29\cdot 10^5$ & $3.39\cdot 10^6$ & & \\\hline
6 & $2.53\cdot 10^7$ & $5.14\cdot 10^8$ & & \\\hline
7 & $2.78\cdot 10^8$ & $1.55\cdot 10^{10}$ & & \\\hline
\end{tabular}
\end{table}


\newpage

\begin{figure}
\caption{ (a) Conformation onto which the se\-quence  
$S_{36}\equiv$ SQKWL\-ERGATRI\-ADGDL\-PVNGT\-YFSC\-KIMENV\-HPLA has been 
designed. In (b) and (c) we display two 
other fully compact target structures, used also as natives
(for another sequences). These conformations were
generated by collapsing a homopolymeric chain at low temperature.}
\end{figure}

\begin{figure}
\caption{Distribution $n_2(\Delta E)$ associated with two 
pointlike
mutations, carried out in three different sequences, with gaps
$\delta=1.3$ (dashed curve), $\delta=1.6$ (continuous curve) and $\delta=2.5$
(dotted curve) respectively. The three ''root'' sequences display no appreciable 
similarity.} 
\end{figure}

\begin{figure}
\caption{Number of sequences which fold into the
conformation shown in Fig.1(a) and obtained from all possible
four-aminoacid pointlike mutations of S$_{36}$ that still preserve the gap.}
\end{figure}

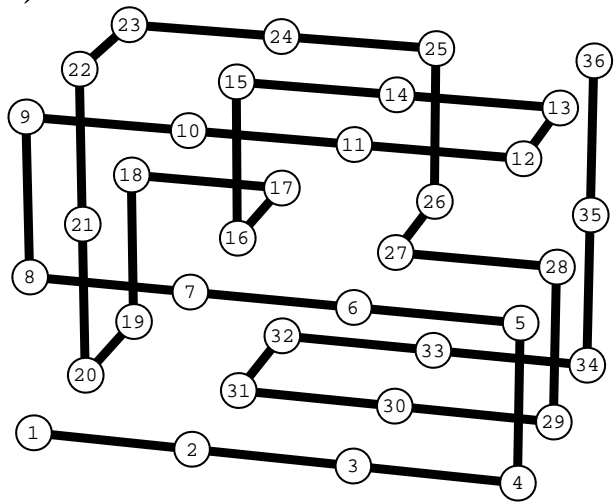
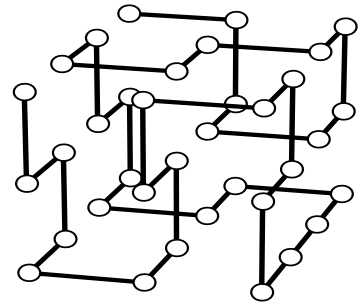
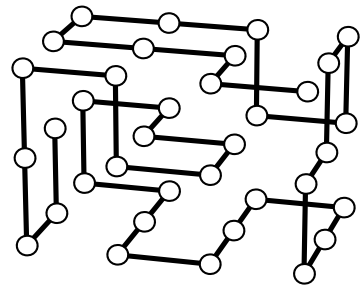
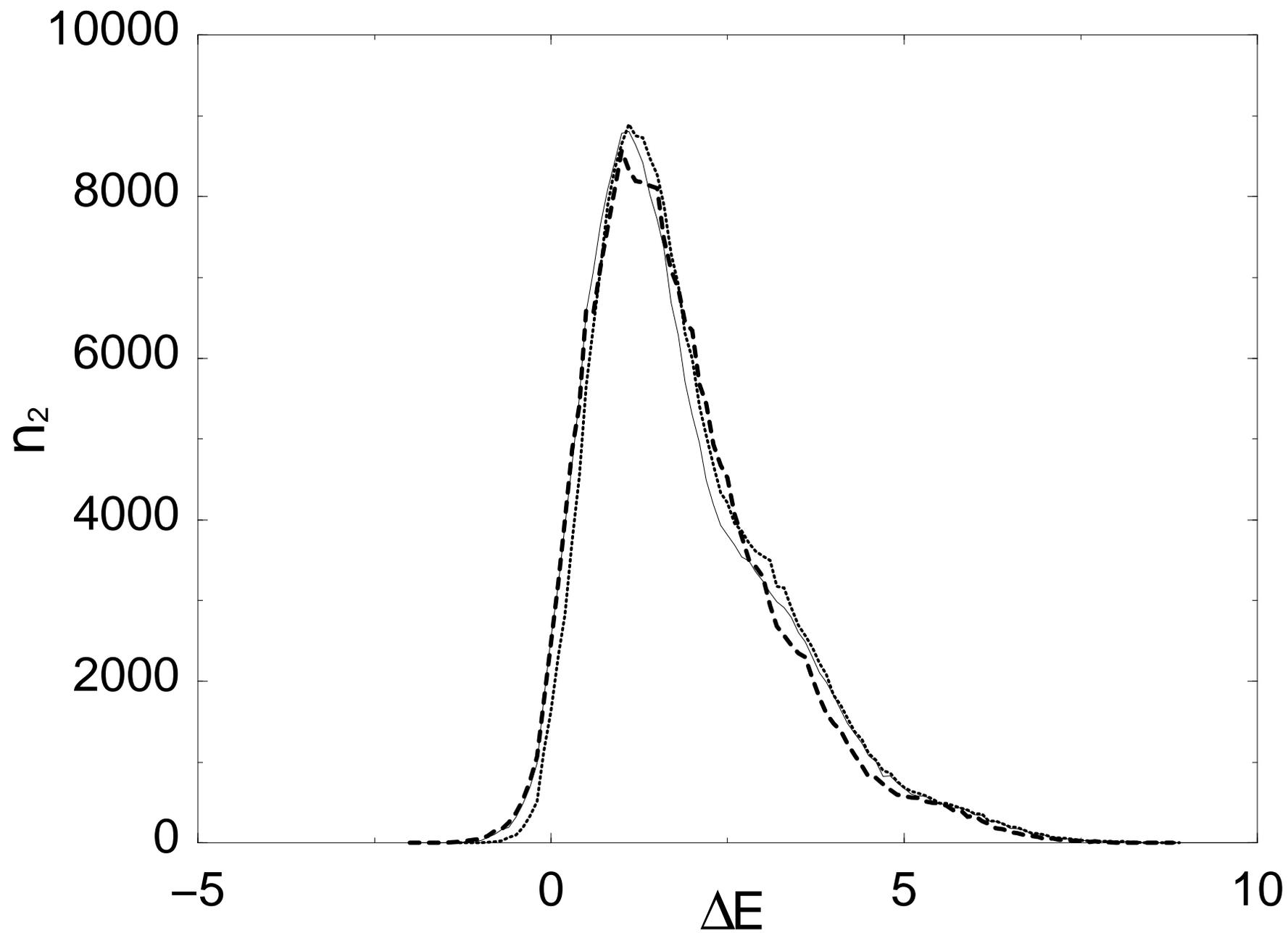
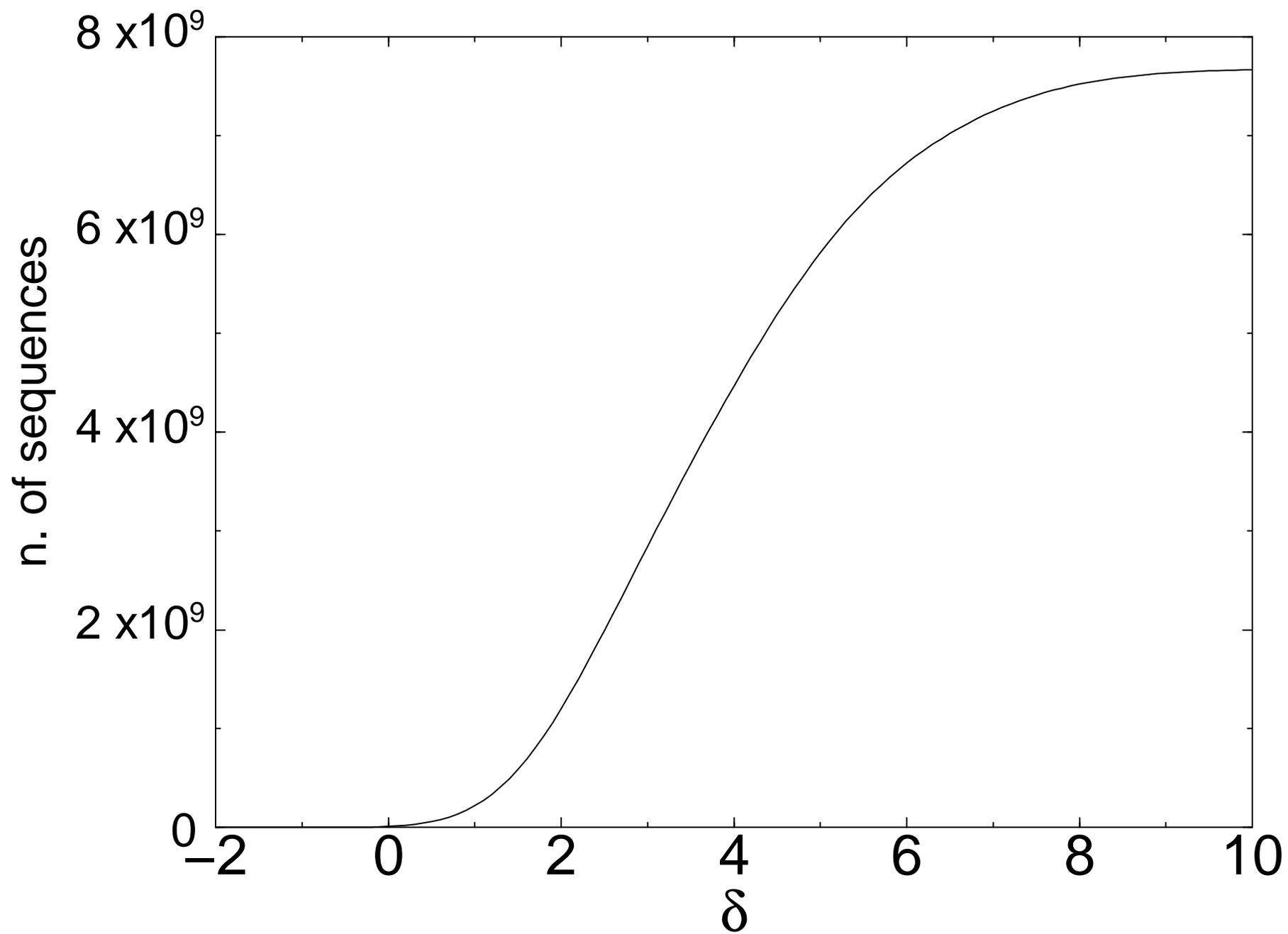
\end{document}